\documentclass[12pt,preprint,longabstract]{aastex}

\slugcomment{Resubmitted: March 20, 2007}

\newcommand{\chase}     {ChASeM33}
\newcommand{\oursnr}{M33SNR\,21}
\newcommand{\oursnrlong}{M33SNR\,21}
\newcommand{\kms}       {\ensuremath{\;\mathrm{km}\;\mathrm{s}^{-1}}}

\newcommand{\cm}        {\ensuremath{\;\mathrm{cm}}}

\newcommand{\pc}        {\ensuremath{\;\mathrm{pc}}}
\newcommand{\kpc}       {\ensuremath{\;\mathrm{kpc}}}
\newcommand{\keV}       {\ensuremath{\;\mathrm{keV}}}
\newcommand{\Kelvin}    {\ensuremath{\;\mathrm{K}}}
\newcommand{\ks}        {\ensuremath{\;\mathrm{ks}}}
\newcommand{\erg}       {\ensuremath{\;\mathrm{erg}}}
\newcommand{\second}    {\ensuremath{\;\mathrm{s}}}
\newcommand{\etal}      {{\it et al.\/}}
\newcommand{\eg}        {{\it e.g.\/}}
\newcommand{\Chandra}   {\textit{Chandra\/}}
\newcommand{\XMM}       {\textit{XMM-Newton\/}}
\newcommand{\ROSAT}     {\textit{ROSAT\/}} 
\newcommand{\FUSE}      {\textit{FUSE\/}}
\newcommand{\Einstein}  {\textit{Einstein\/}}
\newcommand{\Halpha}    {\textrm{H}$\alpha$}
\newcommand{\HII}       {\textrm{H}\,\textsc{ii}}
\newcommand{\fion}[2]   {[\textrm{#1}\,\textsc{#2}]}

\shorttitle{\Chandra{} X-ray Imaging Spectroscopy of \oursnrlong{}}
\shortauthors{Gaetz \etal}

\begin{document}

\title{\Chandra{} ACIS Survey of M33 (ChASeM33): 
X-ray Imaging Spectroscopy of \oursnrlong{},
the brightest X-ray Supernova Remnant in M33}

\author{Terrance J. Gaetz\altaffilmark{1}
\email{gaetz@cfa.harvard.edu}
William P. Blair\altaffilmark{2},
John P. Hughes\altaffilmark{3},
P. Frank Winkler\altaffilmark{4},
Knox S. Long\altaffilmark{5},
Thomas G. Pannuti\altaffilmark{6},
Benjamin Williams\altaffilmark{7},
Richard J. Edgar\altaffilmark{1},
Parviz Ghavamian\altaffilmark{2},
Paul P. Plucinsky\altaffilmark{1},
Manami Sasaki\altaffilmark{1},
Robert P. Kirshner\altaffilmark{1},
Miguel Avillez\altaffilmark{8,9},
and
Dieter Breitschwerdt\altaffilmark{8}}

\altaffiltext{1}{Harvard-Smithsonian Center for Astrophysics,
    60 Garden Street, Cambridge, MA 02138}

\altaffiltext{2}{Department of Physics and Astronomy, \\
Johns Hopkins University, 3400 North Charles Street, Baltimore, MD 21218}

\altaffiltext{3}{Department of Physics and Astronomy, \\
Rutgers University, 136 Frelinghuysen Rd., Piscataway, NJ 08854-8019}

\altaffiltext{4}{Department of Physics, \\
McCardell Bicentennial Hall 526,
Middlebury College, Middlebury, VT 05753}

\altaffiltext{5}{Space Telescope Science Institute, \\
3700 San Martin Drive, Baltimore, MD 21218}

\altaffiltext{6}{Space Science Center, \\
200A Chandler Place, Morehead State University, Morehead, KY 40351}

\altaffiltext{7}{Department of Astronomy, Box 351580, University of 
Washington, Seattle, WA  98195}

\altaffiltext{8}{Institut f\"ur Astronomie, Universit\"at Wien,\\
     T\"urkenschanzstra{\ss}e 17, A-1180 Wien, Austria}

\altaffiltext{9}{Department of Mathematics, University of \'Evora,\\
              R. Rom\~ao Ramalho 59, 7000 \'Evora, Portugal}

\begin{abstract}

We present and interpret new X-ray data for \oursnrlong{}, the brightest X-ray
supernova remnant (SNR) in M33. The SNR is in seen projection against (and
appears to be interacting with) the bright \HII{} region NGC\,592. Data for this
source were obtained as part of the \Chandra{} ACIS Survey of M33 (ChASeM33)
Very Large Project. The nearly on-axis Chandra data resolve the SNR into a $\sim5\arcsec$ diameter (20\,pc at our assumed M33 distance of $817\pm58$ kpc) 
slightly elliptical shell. The shell is brighter in the east, which 
suggests that it is encountering higher density material in that direction.
The optical emission is coextensive with the X-ray shell in the north,
but extends well beyond the X-ray rim in the southwest.  Modeling with X-ray 
spectrum with an absorbed {\tt sedov} model yields a shock
temperature of $0.46^{+0.01}_{-0.02}\,\keV$, an ionization timescale of
$n_\mathrm{e} t = 2.1^{+0.2}_{-0.3} \times 10^{12}\, \mathrm{cm}^{-3}\, 
\mathrm{s}$, and half-solar abundances ($0.45^{+0.12}_{-0.09}$).  Assuming 
Sedov dynamics gives an average preshock H density of 
$1.7\pm 0.3\, \mathrm{cm}^{-3}$.  The dynamical age estimate is 
$6500\pm 600\, \mathrm{yr}$, while the best fit $n_\mathit{e} t$ value 
and derived $n_\mathit{e}$ gives $8200\pm 1700\,\mathrm{yr}$; the weighted 
mean of the age estimates is $7600\pm 600\,\mathrm{yr}$.  We estimate an 
X-ray luminosity (0.25-4.5 keV) of 
$(1.2\pm 0.2) \times 10^{37}\,\mathrm{ergs}\, \mathrm{s}^{-1}$ (absorbed), and
$(1.7\pm 0.3) \times 10^{37}\,\mathrm{ergs}\, \mathrm{s}^{-1}$ (unabsorbed),
in good agreement with the recent \XMM{} determination.  No significant excess 
hard emission was detected; the luminosity 
$\le 1.2\times 10^{35}\, \mathrm{ergs}\, \mathrm{s}^{-1}$ (2-8 keV)
for any hard point source.

\end{abstract}


\section{Introduction}
\label{sec:introduction}

Multiwavelength studies of nearby galaxies are becoming increasingly
effective at providing statistically interesting samples of many
classes of objects, including supernova remnants (SNRs).  At
a distance of 817$\pm$58~kpc \citep{2001ApJ...553...47F}
and with a relatively face-on orientation 
\citep[$i=55\arcdeg\pm1\arcdeg$;][]{1989AJ.....97...97Z},
the late-type Sc spiral galaxy M33 is a key galaxy for
such studies.  At the assumed distance to M33, 1\arcsec\ subtends 4\pc{},
allowing the morphology of some objects to be studied and
many confused or crowded regions to be at least partially
resolved, depending on the wavelength band, instrumentation,
and available spatial resolution.

Nearly 100 SNRs have been identified in M33, based on a combination
of radio and optical imaging and spectroscopy 
\citep{1978A&A....63...63D,
1979A&A....80..212S,
1985ApJ...289..582B,
1986A&AS...64..237V,
1990ApJS...72...61L,
1993ApJ...407..564S,
1998ApJS..117...89G,
1999ApJS..120..247G}.
M33 has also been surveyed by each of the imaging X-ray
missions, including \Einstein{}
\citep{1981ApJ...246L..61L,
1988ApJ...325..531T},
\ROSAT{} \citep{1995ApJ...441..568S,
1996ApJ...466..750L,
2001A&A...373..438H},
and \XMM{}
\citep{2003AN....324...85P,2004A&A...426...11P,2006A&A...448.1247M}.
The \ROSAT{} survey of \citet{1996ApJ...466..750L} found 12 of the 98
optically identified SNRs in the catalog of 
\citet[hereafter GKL98]{1998ApJS..117...89G}, and the \XMM{} survey of
\citet{2003AN....324...85P,2004A&A...426...11P} brought the total
number of X-ray SNR identifications to 21.
\citet{2005AJ....130..539G} used archival {\it Chandra\/}
data on M33 in
comparison with the optical data sets to detect X-ray counterparts
to 22 of the 78 GKL98 SNRs within the \Chandra{}\ fields of view available 
at that time.  Using the characteristics of this X-ray sample, X-ray sources 
without optical or radio counterparts but with X-ray hardness ratios 
similar to those of confirmed SNRs were also identified as candidate SNRs.
This earlier work was largely responsible for motivating the \Chandra{} ACIS 
Survey of M33 (\chase{}), a Very Large Project to examine the 
X-ray point- and extended-source populations and diffuse X-ray emission 
in M33 \citep{2007Sasaki..in..prep}.
At this writing, the survey 
is in progress and nearly complete.  Here we choose to highlight observations 
of a single region in M33 to demonstrate the benefits that this survey 
will ultimately provide over much of the galaxy.

\begin{figure}
\epsscale{0.5}
\plotone{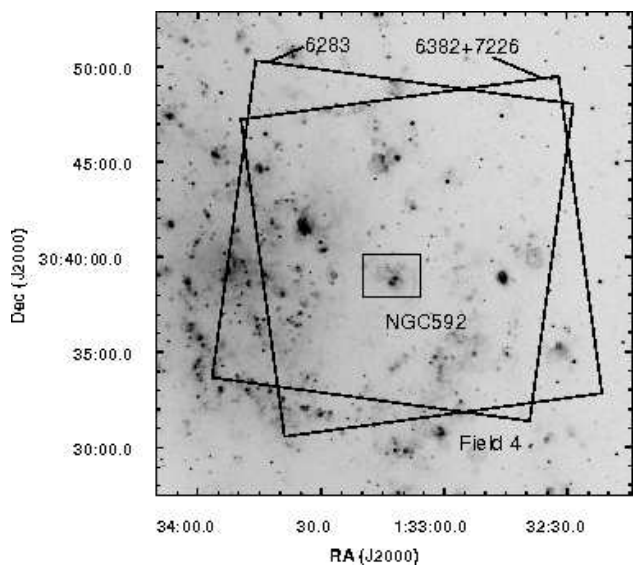}
\caption{\label{fig:fov.overview}
Large squares indicate the $16\farcm8\times16\farcm8$ 
ACIS-I detector footprint for
the pointings closest to the SNR (Field 4), superposed
on a deep \Halpha{} image taken with 
the Burrell Schmidt telescope at Kitt Peak \citep{2006AAS...208.0301M}.  
The small rectangle indicates the region covered by 
Figure~\ref{fig:image.opt.ngc592}
}
\end{figure}

\begin{figure}
\epsscale{0.5}
\plotone{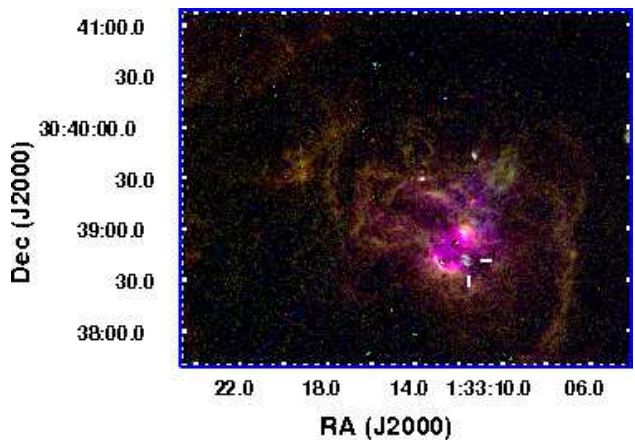}
\caption{\label{fig:image.opt.ngc592}
False color RGB composite of continuum-subtracted images from the 
LGGS of the region surrounding \oursnr{}; 
\Halpha{} is shown in red, \fion{S}{ii} in green, and \fion{O}{iii} in 
blue. The SNR appears in projection against the giant \HII{}
region NGC\,592.  The two bright cores of the \HII{} region
are seen in magenta, and some of the faint outer emission
in the northeast can also be seen.  \oursnr{} appears as the
indicated cyan structure.
}
\end{figure}

\oursnrlong{} (SNR \#21 in the GKL98 catalog; also \#121 in 
\citealt{2004A&A...426...11P} and \#108 in \citealt{2006A&A...448.1247M}) 
is the brightest X-ray SNR in M33 and is located in the outskirts of the 
giant \HII{} region NGC\,592,
which is $\sim$9\arcmin\ due 
west of the galaxy's nucleus (see Fig.~\ref{fig:fov.overview}).
The relation between the SNR and the \HII{} region is shown in 
more detail in Figure~\ref{fig:image.opt.ngc592}, an RGB composite image 
constructed from continuum-subtracted Local Group Galaxy Survey data 
(LGGS\footnote{{\tt http://www.lowell.edu/users/massey/lgsurvey.html}};
\citealt{2006AJ....131.2478M}), 
using \Halpha{} (red), \fion{S}{ii} (green), and \fion{O}{iii} (blue) 
narrow band images; see \S\ref{sec:imaging.analysis} for further details.  
The \HII{} region consists of a pair of bright cores 
separated by $\sim25\arcsec$ which are surrounded by extensive 
($\sim 2\arcmin\times2\arcmin$) faint filamentary structure.  As 
indicated, the SNR stands out by virtue of its elevated [S~II] emission 
relative to H$\alpha$, a signature of shock-heated gas (see, 
\eg{}, \citealt{2004ApJS..155..101B}).
\clearpage
\begin{figure}
\centering
\epsscale{1.0}
\plotone{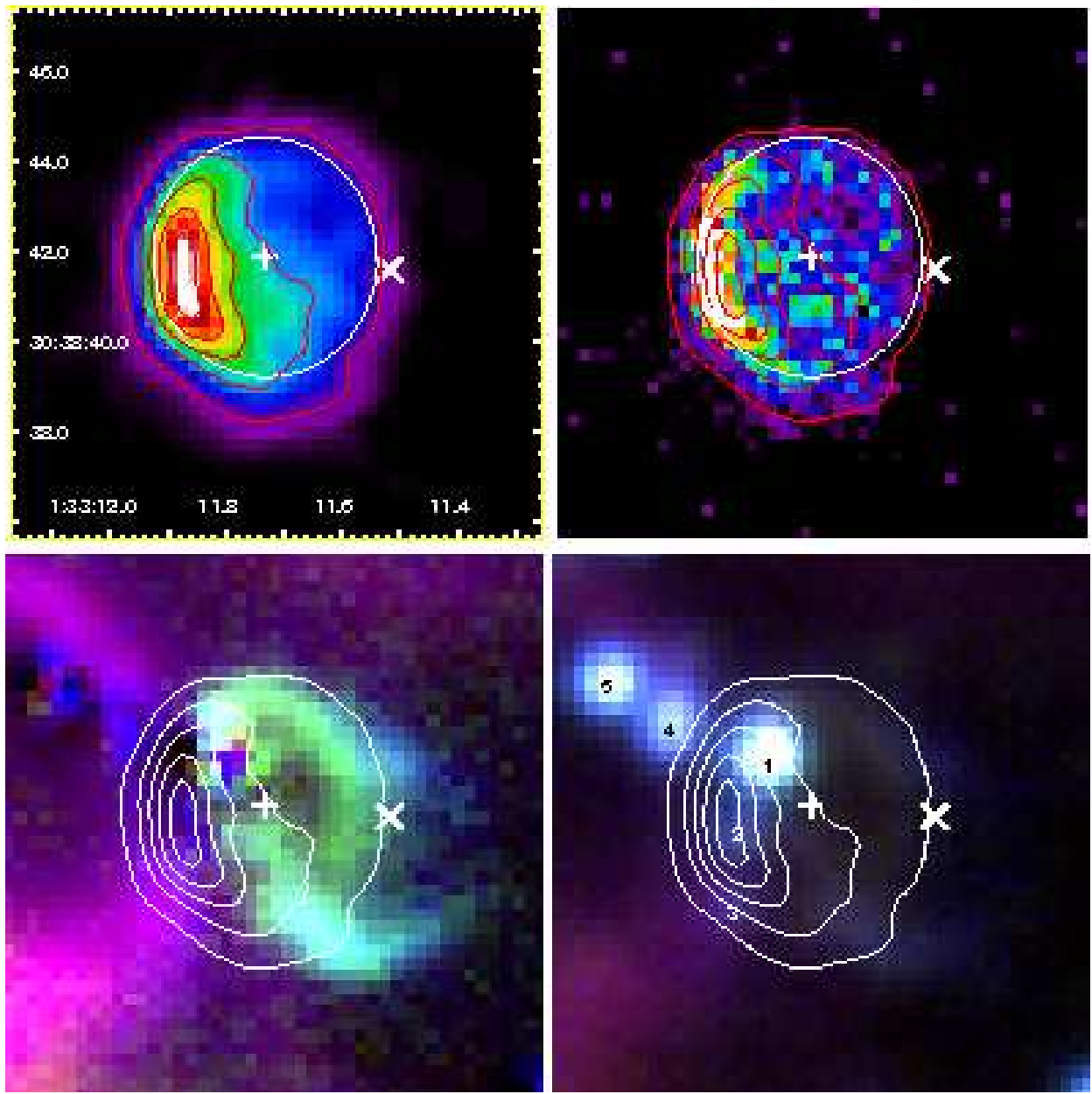}
\caption{\label{fig:image.compare.xray.opt}
Top left: Exposure-corrected (0.35-4\keV{}) \Chandra{} image, binned by 
0.5 ACIS pixels in each direction and smoothed with a 3-bin radius
Gaussian.  The coordinate axes are labeled with J2000 coordinates.
The red contours are from a similarly smoothed 
image for total counts, ranging from 2--18 cts/bin in steps of 4 ct/bin;
the background level is 0.03 ct/bin.  The white ellipse
shows the elliptical shell model dimensions.  The central white ``+''
is the X-ray center, and the white ``$\times$'' to the west is the centroid
position for the nonthermal radio source (see text).
Top right: The result of 10 iterations of the Lucy-Richardson
deconvolution algorithm.  The slight shift of the bright ridge to larger
radii is an artifact.
Bottom panels:  False color RGB composite of the optical images, with
\Halpha{} (red), \fion{S}{ii} (green), \fion{O}{iii} (blue).
Bottom left: Continuum subtracted LGGS images, overlaid with X-ray contours.
Bottom right: LGGS images prior to continuum subtraction, overlaid
with X-ray contours.  The numbers identify stars from 
the LGGS catalog (see text).
}
\end{figure}
\clearpage
\citet{1981ApJ...246L..61L}
first detected \oursnrlong{} in X-rays with \Einstein{}, and suggested
that it was an SNR based on its extremely soft spectrum.  
\citet{1993ApJ...418..743G} confirmed the SNR identification by combining 
radio (VLA and Westerbork), optical (imaging and spectral), and 
X-ray (\ROSAT{}) data.  The nonthermal nature of the radio emission was 
important for confirming the SNR in the midst of otherwise thermal 
emission from the \HII{} region.  \citet{1993ApJ...418..743G} measured 
the density-sensitive optical \fion{S}{ii} lines and found 
$n_\mathrm{e} = 270 \cm^{-3}$, indicating that a dense preshock 
environment is likely responsible for the high X-ray emissivity.  
The high luminosity and dense environment are also an indication that the
SNR is indeed embedded within the \HII{} region and not just seen in 
projection \citep{1993ApJ...418..743G}.  The \ROSAT{} data allowed 
estimates of $\sim$500 $\rm km~s^{-1}$ for the shock velocity and 
$\sim 4 \times 10^{6}\Kelvin{}$ for the postshock temperature.

More recently, \citet{2004A&A...426...11P} observed \oursnrlong{} as part 
of a deep \XMM{} survey of M33 and obtained an absorbed 
0.2--4.5\,keV X-ray flux of 
$(1.41\pm0.02)\times 10^{-13}\,\erg{}\,\second^{-2}
\cm^{-2}$, which implies an absorbed X-ray luminosity of 
$(1.13\pm0.14)\times 10^{37}\,\erg{}\,\second^{-1}$ at our assumed
distance to M33.
The SNR has been cataloged and called by a variety of names, including
M33\,X-3, 2E\,0130.3+3023, 013022+30233, 022+233, GKL\,21, G98-21,
GKL98\,21, GDK\,29, RX\,J0133.1+3038, and XMMU\,J013311.6+303841.

We provide below a detailed analysis of the \chase{} data for \oursnr{}.  
The data reduction and processing steps are described in 
\S\ref{sec:xray.observations.and.data.reduction}.
We select the portion of the data with the best imaging 
resolution and compare with optical data for the region, 
finding very different
morphologies for the optical and X-ray emitting components of
this SNR (\S\ref{sec:imaging.analysis}).  We then combine the
high resolution X-ray data with other portions of the \chase{} data 
(in which the SNR was farther off-axis) to improve the statistics and 
perform X-ray spectral analyses (\S\ref{sec:spectral.analysis}).  
We derive global SNR parameters assuming a Sedov model, and 
compare and contrast \oursnr{} with similar SNRs in the LMC --
N49 and SNR\,0506-68.0 -- in \S\ref{sec:discussion}.  The last 
section, \S\ref{sec:conclusions}, summarizes our results and conclusions.

\section{X-ray Observations and Data Reduction}
\label{sec:xray.observations.and.data.reduction}

The \chase{} survey was designed to cover the inner and most crowded regions
of M33 with seven fields, each with a total exposure of 200\ks{} split 
into two observing intervals; see \citep{2007Sasaki..in..prep}
for further details on the survey.  ACIS-I is the primary detector, 
while ACIS-S2 and S3 
provide additional (though far off-axis) coverage of portions of the 
galaxy adjacent to the primary positions.  At this writing, both 
$\sim$100\ks{} Field 4 pointings closest to the \oursnr{} position 
have been performed (see Fig.~\ref{fig:fov.overview}).  The first 
Field 4 pointing was split into two pieces, ObsIDs 6382 and 7226, 
totaling 97.2\ks{} after screening for high background.  In these 
data, the SNR is $\sim$1.7\arcmin\ 
off-axis on the ACIS-I2 chip and at the same nominal pointing and roll 
(differing by less than 0.1\arcsec\ in both); we merged 
these datasets for 
subsequent analyses.  The second Field 4 pointing, ObsID 6383, 
was not split and totaled 91.4\ks{} after screening; the SNR 
was $\sim$2.1\arcmin\ off-axis on the ACIS-I3 chip.  The Field 4 
observations are all close 
enough to on-axis to attain good spatial resolution.  We simulated 
the \Chandra{} PSF with 
ChaRT\footnote{{\tt http://cxc.harvard.edu/chart/}}
\citep{2003ASPC..295..477C} using the fitted \oursnr{} spectrum 
(see \S\ref{sec:spectral.analysis}) and obtained half-power diameters
of 0.93\arcsec\ and 1.01\arcsec\ for combined ObsIDs 7226+6382 
and ObsID 6383, respectively.

\oursnr{} was also observed in five other pointings at off-axis 
angles $\sim 8\arcmin$--18\arcmin\ (see Table~\ref{tbl:gkl21.obsids}).  
Although the rapid degradation of {\it Chandra's\/} spatial resolution 
at large off-axis angles (\Chandra{} 
POG\footnote{The Chandra Proposers' Observatory Guide, Version 8.0})
makes these observations unsuitable for spatial studies of this SNR, the 
source is bright and isolated enough from other nearby sources to make 
these data useful for investigating the spectrum of the SNR as a whole.
Moreover, ObsIDs 6380 (Field 3) and 6388 (Field 7) imaged the SNR on 
the ACIS-S3 chip which has superior low-energy response and thus 
provides improved information about the softest X-rays.

We reprocessed the X-ray data from the Level 1 files to remove pixel 
randomization and apply time-dependent gain changes.  We screened for 
background flares using the ACIS-I3 light curves except for the 
ACIS-S3 data, for which we used the ACIS-S3 light curves.  We used 
CIAO version 3.3.0.1 and CALDB version 3.2.0.  
In the Field 4 ObsIDs 6282 and 7226, the source dithered across
two ACIS I2 ``bad'' columns (162, 196), resulting in the loss of
$\sim13$\% of the counts from the SNR.  These columns had been marked
bad because of a slight excess background.  However, the SNR spans
only a short stretch of the columns, and a careful examination of the
rest of the data for these columns showed that the spurious background 
would be negligible for the SNR extraction region ($\sim 0.3$ ct for 
column 162, effectively 0 for column 196).  Accordingly, we restored 
those columns for the analysis to improve the statistics.  All the 
Field 4 data were merged for the imaging analysis to improve the 
statistics.  We applied a subpixel event repositioning (SER) correction 
to the data using software developed by \citet[software available from 
the \Chandra{} contributed software 
page\footnote{\tt http://cxc.harvard.edu/cont-soft/soft-exchange.html}]
{2004ApJ...610.1204L} to improve the spatial resolution.  The SER algorithm
uses the distribution of pulse height amplitudes (PHAs) within the
$3\times 3$ pixel event islands to improve the event centroiding.

From the processed event list, we construct a 0.35--4\keV{} ``total
counts'' image (ct/bin) by binning the events from the merged
$\sim200\ks{}$ Field 4 data set by 1/2 ACIS pixel in each
direction ($\sim 1/4\arcsec\times 1/4\arcsec$ bins).  We divide this
counts image by a corresponding exposure map constructed using the fitted 
SNR spectrum (see \S\ref{sec:spectral.analysis}) to generate an exposure
corrected image ($\mathrm{phot}\,\mathrm{cm}^{-2}\,\mathrm{s}^{-1}$).  This
exposure-corrected image is then smoothed with a 3 bin radius Gaussian
kernel.  The image was overlaid by contours from a similarly smoothed
version of the original total counts image, with contours ranging from 2 ct/bin
to 18 ct/bin at intervals of 4 ct/bin.  The corresponding background level is
$\sim 0.03$ ct/bin.
The processed X-ray imaging data are shown and compared with optical images 
of the region in Figure~\ref{fig:image.compare.xray.opt}; see
\S\ref{sec:imaging.analysis} for a discussion of the optical data.
The upper left panel of Figure~\ref{fig:image.compare.xray.opt} shows
the smoothed, exposure-corrected 0.35--4\keV{} image from the combined 
$\sim200\ks{}$ Field 4 data set, binned to 1/2 ACIS pixel in each
direction, with contours from the smoothed total counts image as
described above.
The X-ray imagery reveals a strong asymmetry in the surface brightness 
distribution.  The bright clump in the east has a surface 
brightness $\sim 5$ times that of the fainter portions of the SNR.  
To examine potential finer spatial structure of \oursnr{}, we experimented 
with Lucy-Richardson deconvolutions of this image.  The upper right panel 
of Figure~\ref{fig:image.compare.xray.opt} shows the result for 10 iterations 
with the CIAO tool {\tt arestore} using the combined Field 4 
(ObsIDs 6382, 7226, and 6383) total counts image, and merged ChaRT simulations
for the PSF.  The deconvolution shows an arc of bright emission in the east, 
and possibly some patchy structure within the interior.  The slight shift of 
the bright ridge to larger radii is an artifact of the reconstruction.

\begin{deluxetable}{llccccc}
\tablecaption{
  \Chandra{} X-Ray Observations
  \label{tbl:gkl21.obsids}
}
\tablehead{
     \colhead{Field}
   & \colhead{Obsids}
   & \colhead{$\theta$\tablenotemark{a}}
   & \colhead{Exp}
   & \colhead{CCD}
   & \colhead{Counts\tablenotemark{b}} \\
     \colhead{}
   & \colhead{}
   & \colhead{(arcmin)}
   & \colhead{$(\ks)$}
   & \colhead{}
   & \colhead{(0.35--4\keV)}
}
\startdata
4       
  & 7226, 6382       
  &   1.66       
  &  97.2      
  & I2  
  &  1850 \\

  & 6383             
  &   2.06       
  &  91.4      
  & I3  
  &  1740 \\
5       
  & 6184, 7170, 7171 
  &   7.77       
  & 100.1      
  & I0  
  &  1470 \\
1       
  & 6376             
  &   8.51       
  &  93.1      
  & I1  
  &  1430 \\
3       
  & 6380             
  &  11.26       
  &  89.7      
  & S3  
  &  2155 \\
2       
  & 6378             
  &  16.21       
  & 103.7      
  & S2  
  &  1746 \\
7       
  & 6388             
  &  17.60       
  &  88.7      
  & S3  
  &  2470 \\
\enddata
\tablenotetext{a}{
Off-axis angle from the optical axis.}
\tablenotetext{b}{Estimated background-subtracted counts (0.35-4\keV)}
\end{deluxetable}
\begin{figure}[t]
\epsscale{0.5}
\plotone{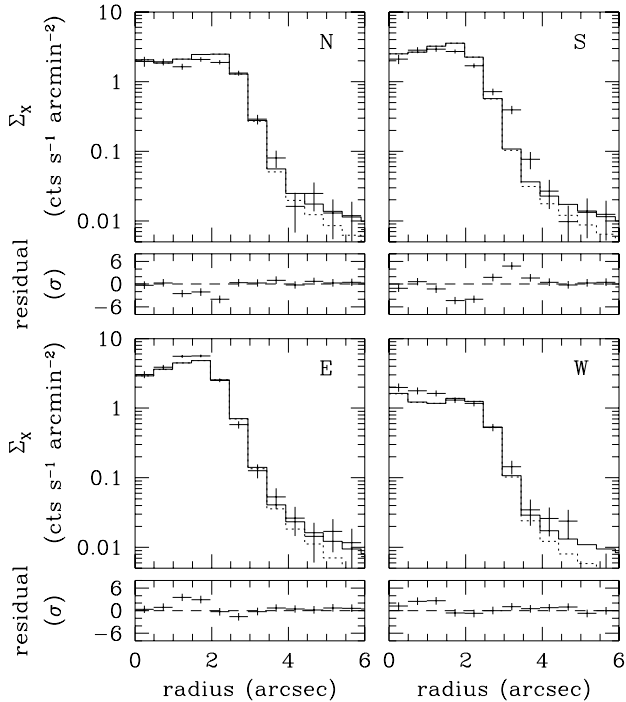}
\caption{\label{fig:elliptical.shell.model}
Elliptical shell fit to Field 4 image data.
The panels compare the data and projected model for the four quadrants
centered on North, South, East, and West.
}
\end{figure}
\section{Imaging Analysis}
\label{sec:imaging.analysis}

For optical images of the \oursnr{} field, we took data from the Local
Group Galaxies Survey (LGGS; \citealt{2006AJ....131.2478M}) which used 
the KPNO 4m telescope and Mosaic CCD camera to survey most of M33  
(3 overlapping fields) through narrow-band \Halpha{}, \fion{O}{iii} 
$\lambda$\ 5007, and \fion{S}{ii} $\lambda\lambda 6716, 6731$\ filters, plus 
broadband \textit{UBVRI}.  \oursnr{} appears in both the north and central 
fields.  From each Mosaic image we clipped out a small section centered 
on the SNR and precisely aligned these using several dozen field stars.  
We then selected the images with the best seeing in each of the 3 emission 
lines (all from the north field) and in the {\it V\/} and {\it R\/} 
bands (both from the central field) and matched the point-spread 
functions.  Finally we scaled the broadband images appropriately, 
and performed continuum subtraction to remove the stars as effectively as
possible.  In this case, we subtracted \textit{V} from \fion{O}{iii}, and \textit{R} from 
both \Halpha{} and \fion{S}{ii}.  The result is shown 
in Figure~\ref{fig:image.opt.ngc592}.  
The lower left panel of Figure~\ref{fig:image.compare.xray.opt}
is an enlarged version of the same image with colors rescaled to emphasize the 
SNR, and the lower right panel is an RGB composite of the narrow-band 
images prior to continuum subtraction to show the locations of the bright 
stars.  X-ray contours have been plotted on both to facilitate comparison 
with the upper panels.  Based on the \textit{V} magnitudes and 
\textit{UBVRI} colors obtained by the LGGS, the three stars seen in 
projection (labeled ``1'', ``2'', and ``3'') and the star furthest 
to the northeast (``5'') are O stars in M33.  
Typical O star X-ray luminosities are  $\la 10^{33} \erg{}\,\second^{-1}$
\citep{1981ApJ...245..163V} and an O star would be undetectable in the 
500\ks{} of data used in our analysis.  The ``star'' closest to the northeast rim 
, ``4'', consists of three stars, two of which appear to be O stars.  The third
(much fainter) star has an unusual color, but no X-ray events were detected 
in the 190\ks{} of the Field 4 imaging data 
(Fig.~\ref{fig:image.compare.xray.opt}).  
The X-ray data are not significantly contaminated by stellar emission.
\begin{figure}
\epsscale{0.5}
\plotone{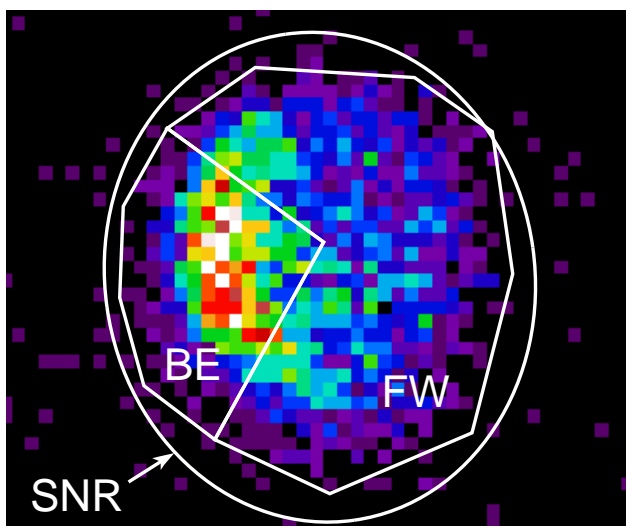}
\caption{\label{fig:xray.wedge.reg}
Unsmoothed total counts image, binned to 1/2 ACIS pixels.
The Field 4 Bright Eastern (``BE'') and Faint Western (``FW'') wedge extraction 
regions are indicated. The extraction region for the full SNR is 
indicated by the ellipse labeled ``SNR''.
}
\end{figure}

To characterize the overall shape of the X-ray SNR, we fitted the 0.35-4\keV{} 
image data with an elliptical shell model (assuming that the symmetry 
axis lies in the plane of the sky), with density within the 
shell varying azimuthally as 
\begin{equation}
\label{eqn:elliptical.shell.model}
n(\theta) \propto n_0 (1 + 0.5 A (1 + \cos(\theta-\theta_0))) .
\end{equation}
Figure~\ref{fig:elliptical.shell.model} shows four projections of the model 
onto the data (including the effect of the simulated \Chandra{} PSF) for 
quadrants centered on north, south, east, and west.  The greater brightness 
in the eastern quadrant compared to the western quadrant is evident.
The model shows the SNR to be slightly elliptical (axis ratio 1.07) with 
the position angle of the major axis $\sim 9^{\circ}$ east from north.  
The semimajor (semiminor) axis is 2.65\arcsec (2.48\arcsec) with an
uncertainty of $\sim \pm 0.05\arcsec$.  At the assumed distance of M33, 
this corresponds to a semimajor (semiminor) axis of 10.5\pc{} (9.8\pc{}) with
an uncertainty of $\sim\pm0.2\,\mathrm{pc}$, 
and an overall size uncertainty of $\sim 0.7\,\mathrm{pc}$ when the
distance uncertainty is included.  The surface brightness peaks at 
position angle $\sim 104\degr$ east from north.  The fitted value for 
the factor $A$ in the elliptical shell model is 1.19; this implies the
ambient density on the bright side is about a factor of $1+A \approx 2$ 
higher than that on the faint side.  
The outline of the elliptical
shell model is indicated by an ellipse in the upper panels of
Figure~\ref{fig:image.compare.xray.opt}.
Compared to the fitted ellipse, the X-ray emission extends slightly further 
to the south than other directions (Fig.~\ref{fig:image.compare.xray.opt}, 
upper left panel; see also the comparison between the northern vs. southern 
quadrants in Figure~\ref{fig:elliptical.shell.model}).  
The fractional thickness of the shell,
$\Delta R_\mathit{s}/R_\mathit{s}$, is poorly constrained because 
the SNR is only barely resolved; we found that a value 
of $0.065$ provides an adequate description for the 
limb-brightening of the SNR's shell.  
The resulting center position is 
$(\alpha_\mathrm{J2000},
\delta_\mathrm{J2000})$=($01^\mathrm{h}33^\mathrm{m}11\fs73$, 
+$30\arcdeg38\arcmin41\farcs9$), indicated 
by the central ``+'' in each panel of 
Figure~\ref{fig:image.compare.xray.opt}.  This position agrees well
with the \XMM{} position determination \citep{2006A&A...448.1247M} 
of $(\alpha_\mathrm{J2000}, \delta_\mathrm{J2000})
  = (01^\mathrm{h}33^\mathrm{m}11\fs76$,
+$30\arcdeg38\arcmin42\farcs1)$, well within the \XMM{} uncertainty of 
0.55\arcsec.

\begin{figure}[t]
\epsscale{0.5}
\includegraphics[width=0.4\textwidth,angle=270,clip]
  {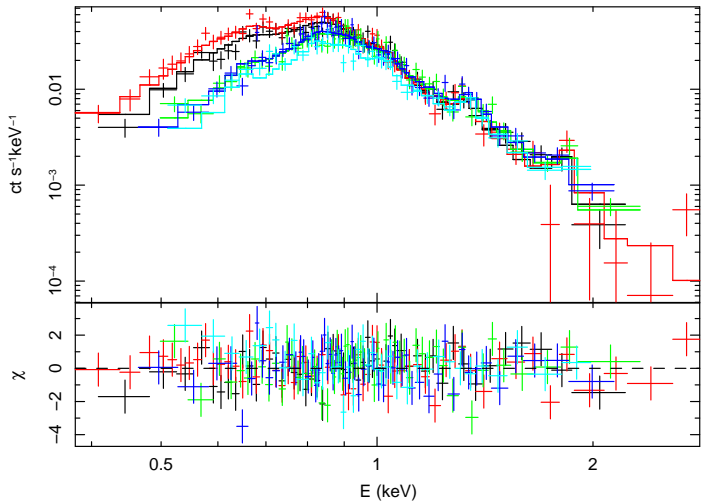}
\caption{\label{fig:spectrum.sedov.solarabd}
\Chandra{} ACIS-I and ACIS-S3 spectra and best fit {\tt sedov} model 
of \oursnr{}; the curves which are higher at low energies are ACIS-S3,
and the rest are ACIS-I data.
}
\end{figure}

Using techniques similar to those in the \citet{2006Book...W} review
paper, we also examined the merged Field 4 hard band (3--8\keV{}) image for any 
indications of a point source or emission from a plerion.  The extraction 
region for the whole SNR includes only 14 counts in the 3--8\keV{} band during 
the entire 190\,ks integration.  The best fit {\tt XSPEC sedov} model (see
\S\ref{sec:spectral.analysis}) predicts $11\pm5$ counts in this band, 
so there is no evidence for a significant excess of hard emission.  
The brightest potential point source has a total of 3 counts 
(3--8\keV{}) within two adjacent pixels, providing a 3$\sigma$ count rate upper
limit of $\la 4.2\times 10^{-5}\,\mathrm{ct}\,\mathrm{s}^{-1}$.  
If we assume a Crab-like powerlaw spectrum (powerlaw index $\alpha=2.05$), 
the same absorbing column as for the best fit {\tt sedov} model
and the response for the ACIS-I2 observation, this count rate limit allows 
an unabsorbed flux upper limit for a point source of 
$\la 1.5\times 10^{-15}\,\mathrm{erg}\,\mathrm{cm}^{-2}\,\mathrm{s}^{-1}$
(2--8\,\keV{}) corresponding to an unabsorbed X-ray luminosity upper 
limit of $\la 1.2\times 10^{35}\,\mathrm{erg}\,\mathrm{s}^{-1}$ for
the same band.

The centroid of the nonthermal radio source 
\citep[precessed to J2000]{1993ApJ...418..743G}
is 2.8\arcsec\ west of the center of the X-ray emission, and indicated
by a ``$\times$'' toward the right in each panel of 
Figure~\ref{fig:image.compare.xray.opt}.
Thermal emission from the \HII{} region may confuse the radio picture 
\citep{1993ApJ...418..743G}.  The resolution of the radio data is 
$\sim 7\arcsec$ so the $\sim 3\arcsec$
offset of the radio centroid from the X-ray center is unlikely to be
significant.

The relative morphologies of the optical and X-ray emissions from \oursnr{}
are interesting and somewhat perplexing. The X-ray SNR appears nearly 
circular, although with the strong brightness gradient already described, 
but the optical SNR appears somewhat more structured and extended.  The 
bright optical emission agrees well with the outer X-ray contour shown in 
Figure~\ref{fig:image.compare.xray.opt} along the northern and 
northwestern X-ray limbs, apparently validating the relative astrometry 
used to align the images.  However, the bright eastern X-ray limb is almost 
devoid of optical emission (although this comparison is hampered by the stellar 
emission) and faint optical emission extends farther to the southwest
from the X-ray contour.  Taken on its own, the optical morphology in the 
southwest is reminiscent of a ``blow-out,'' which is consistent with the 
idea that the density is lower on the side of the SNR away from the bright 
\HII{} cores.  The absence of optical emission on the eastern X-ray peak does 
not appear to be due to obscuration, as the $N_\mathrm{H}$ column derived 
below is not extreme.  

\clearpage
\begin{deluxetable}{llllll}
\tablecaption{
  Spectral Fit Results for M33SNR\,21.
  \label{tbl:gkl21.fits}
}
\tablehead{
     \colhead{Parameter}
   & \colhead{Units}\hspace{3.0cm} 
   & \colhead{BE} 
   & \colhead{FW} 
   & \colhead{SNR} 
   & \colhead{SNR} \\
   & 
   & {F4 data} 
   & {F4 data} 
   & {F4 data} 
   & {F1,3,4,7 data}
}    
\startdata
\cutinhead{const(phabs(vphabs(pshock))) model}
Abundance 
       & (solar\tablenotemark{a})
       & $\displaystyle 0.4^{+0.6}_{-0.2}$
       & $\displaystyle 0.4^{+0.3}_{-0.1}$ 
       & $\displaystyle 0.3\pm0.1$
       & $\displaystyle 0.28^{+0.06}_{-0.05}$ \\[5pt]
$kT_\mathit{e}$ 
       & [keV]           
       & $\displaystyle 0.55^{+0.04}_{-0.07}$
       & $\displaystyle 0.61^{+0.04}_{-0.05}$ 
       & $\displaystyle 0.58\pm0.03$
       & $\displaystyle 0.58\pm0.02$  \\[5pt]
$\tau$ 
       & [10$^{12}\,\cm^{-3}\,\sec$]    
       & $\displaystyle 0.7^{+0.4}_{-0.2}$
       & $\displaystyle 0.8^{+0.5}_{-0.3}$
       & $\displaystyle 0.8^{+0.3}_{-0.2} $
       & $\displaystyle 0.6\pm0.1$ \\[5pt]
$N_{\rm H,M33}$\tablenotemark{b} 
       & [10$^{20}\,\cm^{-2}$] 
       & $\displaystyle \le 7.8 $
       & $\displaystyle 2.4^{+14.3}_{-2.0} $
       & $\displaystyle \le 3.0 $
       & $\displaystyle \le 2.9 $            \\[5pt]
$K$\tablenotemark{c} 
       & [$10^{-4} \cm^{-5}$] 
       & $\displaystyle 0.54\pm0.02$
       & $\displaystyle 0.61^{+0.03}_{-0.02}$
       & $\displaystyle 1.36^{+0.03}_{-0.04} $
       & $\displaystyle 1.46^{+0.03}_{-0.02} $ \\[5pt]
$\chi^{2}_\mathit{red}$ (d.o.f.) 
       &
       & $\displaystyle 1.11 (51)$
       & $\displaystyle 1.11 (62)$
       & $\displaystyle 1.14 (111)$
       & $\displaystyle 1.14 (259)$ \\[5pt]
\cutinhead{const(phabs(vphabs(sedov))) model}
Abundance 
       & (solar\tablenotemark{a})
       & $\displaystyle 0.5^{+0.6}_{-0.2}$ 
       & $\displaystyle 0.7^{+2.1}_{-0.3}$ 
       & $\displaystyle 0.5^{+0.3}_{-0.2}$ 
       & $\displaystyle 0.45^{+0.12}_{-0.09}$  \\[5pt] %
$kT_\mathit{e}$ 
       & [keV]           
       & $\displaystyle 0.43^{+0.03}_{-0.05}$ 
       & $\displaystyle 0.48^{+0.04}_{-0.08}$ 
       & $\displaystyle 0.46\pm0.02$ 
       & $\displaystyle 0.46^{+0.01}_{-0.02}$  \\[5pt]
$\tau$ 
       & [10$^{12}\,\cm^{-3}\,\sec$]    
       & $\displaystyle 2.4^{+1.5}_{-0.8}$
       & $\displaystyle 2.4^{+1.7}_{-1.2}$
       & $\displaystyle 2.6^{+0.9}_{-0.6}$
       & $\displaystyle 2.1^{+0.2}_{-0.3}$ \\[5pt]  %
$N_{\rm H,M33}$\tablenotemark{b} 
       & [10$^{20}\,\cm^{-2}$] 
       & $\displaystyle \le 9.6$
       & $\displaystyle 4.1^{+17.5}_{-4.1}$
       & $\displaystyle \le 4.0$
       & $\displaystyle \le 3.4$            \\[5pt]%
$K$\tablenotemark{c} 
       & [$10^{-4}\,\cm^{-5}$] 
       & $\displaystyle 0.51\pm0.02$
       & $\displaystyle 0.49\pm0.02$
       & $\displaystyle 1.08\pm0.03$
       & $\displaystyle 1.20\pm0.02$ \\[5pt] %
$\chi^{2}_\mathit{red}$ (d.o.f.) 
       &
       & $\displaystyle 1.02 (51)$
       & $\displaystyle 1.05 (62)$
       & $\displaystyle 1.07 (111)$
       & $\displaystyle 1.08 (261)$
\enddata
\tablenotetext{a}{\citet{1989GeCoA..53..197A}}
\tablenotetext{b}{N$_{\rm H,MW}$ was fixed at $5.5 \times 10^{20}\,\cm^{-2}$.
                 The abundances for $N_{\rm H,M33}$ were fixed at 0.5
                 solar.}
\tablenotetext{c}{Normalization $K=(10^{-14}/(4\pi D^2)) \int
  n_\mathit{e} n_\mathrm{H}\,dV$ where
$D$ is the source distance (in cm), $n_\mathit{e}$ is the electron
number density (cm$^{-3}$) and $n_\mathrm{H}$ is the hydrogen number
density (cm$^{-3}$).
}
\tablecomments{Error ranges are 90\% confidence intervals ($\Delta\chi^2
\le 2.706$ for one parameter).}
\end{deluxetable}
\begin{deluxetable}{lllll}
\tabletypesize{\small}
\tablewidth{0pt}
\tablecaption{
  X-ray fluxes and luminosities based on the {\tt sedov} model fits
  \label{tbl:flux.lum}
}
\tablehead{
     \colhead{Band}
   & \colhead{$F$ (absorbed)}
   & \colhead{$F$ (unabsorbed)}
   & \colhead{$L_X$ (absorbed)}
   & \colhead{$L_X$ (unabsorbed)} \\

   & $(10^{-13}\,\erg\,\mathrm{s}^{-1}\,\cm^{-2})$
   & $(10^{-13}\,\erg\,\mathrm{s}^{-1}\,\cm^{-2})$
   & $(10^{37}\,\erg\,\mathrm{s}^{-1})$
   & $(10^{37}\,\erg\,\mathrm{s}^{-1})$
}
\startdata
0.35--3.0\,\keV
       & $1.46\pm0.03$
       & $1.95\pm0.03$       %
       & $1.2\pm0.2$
       & $1.6\pm0.3$ \\    %
0.245--4.5\,\keV
       & $1.50\pm0.03$
       & $2.15\pm0.04$       %
       & $1.2\pm0.2$
       & $1.7\pm0.3$       %
\enddata
\end{deluxetable}
\section{Spectral Analyses}
\label{sec:spectral.analysis}

The combined Field 4 data (the only X-ray data for which the SNR 
is resolved well enough to extract spectra from individual subregions) has 
$\sim 3600$ counts.  We extracted spectra from two wedge-shaped regions, 
each containing roughly half the total counts:  the bright eastern part 
(``BE'', $\sim 1600$ counts), and the fainter western part 
(``FW'', $\sim 2000$ counts); see Figure~\ref{fig:xray.wedge.reg}.
The BE region has $28$\% of the area.  We also extracted spectra for 
the SNR as a whole using an elliptical region (``SNR''; see 
Figure~\ref{fig:xray.wedge.reg}).  Because ObsIDs 6382 and 7226 fell on the 
ACIS-I2 chip, while ObsID 6383 fell on ACIS-I3, we extracted separate spectra 
for the merged 6382+7226 data and the 6383 data.  The data were grouped 
to a minimum of 25 counts/bin, and the $\chi^2$ statistic was used for 
the fits.  Background spectra were extracted from larger adjacent regions.
In order to improve the fit statistics for the spectrum for the SNR as a whole,
we also extracted spectra for a number of the far off-axis observations:
Field 1 (ACIS-I1), and Fields 3 and 7 (ACIS-S3).  The source and background 
extraction regions were enlarged appropriately to account for the expansion 
of the \Chandra{} PSF at large off-axis angles.  This provided a total of 
$\sim 10000$ counts for the SNR as a whole.  For the ACIS-S3 data 
(Fields 3 and 7), the SNR was imaged near the edge of the chip and 
$\sim 10\%$ of the flux spilled off the chip.

We fitted the spectra using {\tt XSPEC} version 11.3.2.  In each case, we included 
absorption ({\tt phabs}) corresponding to a Galactic column of 
$N_\mathrm{H} = 5.5\times 10^{20}\,\cm^{-2}$ toward M33 
\citep{1995yCat.8028....0S}; this component was frozen during the fits.  
To account for absorption within M33, we included a second absorption 
component ({\tt vphabs}) which was allowed to vary:  the abundances of the 
{\tt vphabs} component were fixed to be 0.5 solar, where we assume 
\citet{1989GeCoA..53..197A} values for the solar abundance set.  The M33 abundance 
at the galactocentric radius of \oursnr{} is 
$\sim 0.5\,Z_\odot$ \citep{1995ApJ...438..170H}.
For the thermal component, we examined collisional ionization equilibrium 
models (which provided very poor fits), and a number of nonequilibrium 
ionization (NEI) models: {\tt nei}, {\tt pshock}, and {\tt sedov}.  
The {\tt nei} and {\tt pshock} models represent impulsive heating to 
a constant temperature, $T$; in the former model, the spectrum evaluated 
at a single ionization timescale, $\tau \equiv n_\mathit{e} t$, while 
in the latter, the spectrum is integrated over ionization timescales from 
$\tau = 0$ to $n_\mathit{e} t$.  In the {\tt sedov} model, the 
nonequilibrium ionization spectrum is integrated over a Sedov-stage 
SNR profile, and the ionization timescale is $\tau = n_\mathit{e,s}\,t_0$ 
where $n_\mathit{e,s}$ is the postshock electron density and $t_0$ is 
the age of the SNR.  See \citet{2001ApJ...548..820B} for a more detailed 
description of the models.  

We performed fits for regions BE, FW, and the whole SNR using the high 
spatial resolution Field 4 data, and also for the whole SNR including data 
from Fields 1, 3, and 7 in addition to Field 4.  The spectra were fitted 
simultaneously with the model parameters (including model normalizations) 
linked.  To account for the loss of flux off the chip for the ACIS-S3 
data, an additional multiplicative {\tt const} model was applied, allowed 
to be free for the ACIS-S3 datasets, and fixed at 1.0 for the other 
datasets.  The resulting {\tt const} fit values were $\sim 0.9$ for 
the ACIS-S3 data, as expected.

All three models had comparable $\chi^2_\mathit{red} \sim 1$, but the 
{\tt sedov} models consistently produced better fits: 1.08 for 261 degrees 
of freedom (dof) for {\tt sedov} vs. 1.14 (259 dof) and 1.21 (259 dof)
for {\tt pshock} and {\tt nei}, respectively.  The {\tt nei} and 
{\tt pshock} models gave lower abundance estimates ($0.15$--$0.19 Z_\odot$ 
and $0.22$--$0.34 Z_\odot$, respectively) compared to the {\tt sedov} model
($0.36$--$0.57 Z_\odot$).  The temperatures were 
$0.52$--$0.56\keV{}$ ({\tt nei}), $0.56$--$0.6\keV{}$ 
({\tt pshock}), and $0.44$--$0.47\keV{}$ ({\tt sedov}).  The {\tt pshock} 
and {\tt sedov} fit parameters are listed in Table~\ref{tbl:gkl21.fits}; 
the listed parameter uncertainties are 90\% confidence intervals 
($\Delta \chi^2 \le 2.706$ for one parameter) based on the statistical errors.

In the {\tt sedov} model, the temperature parameter is the postshock (electron) 
temperature.  In the Sedov self-similar solution the temperature increases (and 
the density decreases) radially inwards so that the emission-weighted 
X-ray temperature is $\sim 1.3$ times the shock temperature 
\citep{1974ApJ...194..329R}. In the other two models, {\tt nei} 
and {\tt pshock}, the temperature parameter is the (constant)
postshock temperature.  If the temperature in the SNR is in fact 
increasing radially inwards as in the Sedov solution, the fitted 
temperature parameters for the {\tt nei} and {\tt pshock} models would 
reflect an average X-ray temperature higher than the temperature at the
shock front; this would be consistent with the generally higher 
temperature parameters in those fits compared to the {\tt sedov} model 
fits (Table~\ref{tbl:gkl21.fits}).

Based on the best fitting {\tt sedov} model, we evaluated the X-ray flux and 
luminosity (absorbed and unabsorbed) for 0.35--3\,\keV{} (the range best 
constrained by the data) and a broader 0.25--4.5\,\keV{} range; the results 
are presented in Table~\ref{tbl:flux.lum}.  The quoted errors include the 
normalization uncertainty (Table~\ref{tbl:gkl21.fits}) and the distance 
uncertainty.  These luminosities may be compared to the luminosities 
(scaled to 817\kpc{}) estimated with \Einstein{} IPC 
($1.7\times 10^{37}\,\mathrm{erg}\,\mathrm{s}^{-1}$ absorbed, 0.15--4.5\,keV; 
\citealt{1981ApJ...246L..61L}), \ROSAT{} PSPC (flux 
$1.4\times 10^{-13}\,\mathrm{erg}\,\mathrm{cm}^{-2}\,\mathrm{s}^{-1}$
implying $L_X = 1.1\times 10^{37}\,\mathrm{erg}\,\mathrm{s}^{-1}$
absorbed, 0.1--2.4\,keV; \citealt{2001A&A...373..438H})
and \XMM{} EPIC (flux $(1.41\pm0.02)\times 
10^{-13}\,\mathrm{erg}\,\mathrm{cm}^{-2}\,\mathrm{s}^{-1}$ 
implying 
$L_\mathit{X} = (1.13\pm0.16)\times 10^{37}\,\mathrm{erg}\,\mathrm{s}^{-1}$
absorbed, 0.2--4.5\,keV; \citealt{2004A&A...426...11P}).  Our luminosity
estimate of
$L_\mathit{X} = (1.2\pm0.2)\times 10^{37}\,\mathrm{erg}\,\mathrm{s}^{-1}$ 
(absorbed, 0.25--4.5\,keV) is in good agreement with the \ROSAT{} 
and \XMM{} estimates and in reasonable agreement with the \Einstein{} estimate.
It is also worth noting that the SNR X-ray luminosity function for M33
does not reach as high as in other nearby galaxies
\citep{2005AJ....130..539G}.  \oursnr{}, the brightest SNR in M33, 
would be only the fourth or fifth brightest SNR if it were in the
Large Magellanic Cloud (LMC), based on the luminosities
reported in \citet{1981ApJ...248..925L}.

\section{Discussion}
\label{sec:discussion}

\subsection{Association with NGC\,592}
\label{subsec:assoc.with.ngc.592}

As noted in \S\ref{sec:introduction}, the SNR lies along the line of
sight to the bright \HII{} region NGC\,592.  The high X-ray luminosity of 
the SNR and the relatively high density interstellar medium (ISM)
inferred from the optical \fion{S}{ii} emission imply that
the SNR is in fact embedded in the \HII{} region \citep{1993ApJ...418..743G}. 
The fact that the X-ray emission from \oursnr{} is brightest on the side toward 
the bright \HII{} region core further supports the association with
NGC\,592.
This suggests that the \oursnr{} progenitor could have been a massive 
star associated with the burst of star formation which produced NGC\,592.

\citet{2006AJ....131..849P} recently reported on \FUSE{} observations of
NGC\,592 in which the \FUSE{} 30\arcsec\ aperture covered the bright
\Halpha{} cores (but excluded \oursnr{}).  The line profiles show extended 
blue absorption wings, indicating the presence of evolved O stars.  Her 
fits to synthetic spectra based on stellar population models are consistent 
with a $4.0\pm0.5$ Myr population at solar metallicity; the FUV models 
are consistent with a range of $\sim 0.4$--1.2 solar.  
\citet{2002MNRAS.329..481B} estimated a population age of 
$\ga4.5$\,Myr, but \citeauthor{2006AJ....131..849P} argues that this 
is inconsistent with the strength of the P-Cygni features.  If we 
assume that the \oursnr{} progenitor belongs to the same population, the
models of \citet{1990A&AS...84..139M} suggest an initial stellar mass
of $\sim 50\,M_\odot$ for half-solar to solar abundances.  Our X-ray 
spectroscopic analysis of \oursnr{} with \Chandra{} data also suggests roughly 
half-solar abundances.  The lack of an X-ray signature from ejecta is likely 
a combination of the age of the SNR ($\sim 5$--10\,kyr) and the limited 
\Chandra{} spatial resolution at the distance of M33.  For the
integrated spectrum for the SNR as a whole, most of the X-ray 
emission arises from the swept-up interstellar medium in M33.  
These high spatial resolution X-ray observations 
apparently do not provide sufficient resolution and statistics to 
separate out high abundance clumps if they are present.

\citet{2006AJ....131..849P} also obtained an intrinsic extinction of 
$E(B-V)_i = 0.07 \pm 0.02$ for the \HII{} region.  For this extinction, 
the LMC interstellar extinction curve \citep{1986AJ.....92.1068F} implies
$N_\mathrm{H} \approx 1.7\times 10^{21}\,\cm^{-2}$, while the Galactic 
extinction curve \citep{1978ApJ...224..132B} implies 
$N_\mathrm{H} \approx 4\times 10^{20}\,\cm^{-2}$.  The analysis of 
the \Chandra{} X-ray spectra provides primarily upper limits 
$N_\mathrm{H} \le (2$--10$)\times 10^{20}\,\cm^ {-2}$ for the M33 
column, consistent with a solar metallicity extinction curve.  However, 
given the complex environment, the position of \oursnr{} along the line of 
sight to NGC\,592 cannot be determined at present.  If the SNR is 
toward the near side of the \HII{} region, the low column for the 
X-ray data could also be consistent with the FUV data and a half-solar 
metallicity extinction curve.

\subsection{Sedov Analysis}
\label{subsec:sedov.analysis}

In the following, we examine the global properties of \oursnr{} based on the
assumption of a Sedov-stage SNR.  \citet{1999A&A...344..295H} investigated 
the properties of nonspherical SNRs using approximate analytic models, 
including the cases of an SNR in an exponential density gradient, and 
an SNR in an off-center $r^{-2}$ powerlaw distribution (for example, an 
explosion in the wind cavity of another star).  The integrated
spectral properties of the SNRs were found to be approximately 
like those of Sedov models, even though the detailed properties are 
expected to vary spatially across 
the SNRs.  The surface brightness ($\propto n^2$) varies more strongly 
with density around the remnant than the temperature ($\propto n$), and the shock 
velocity varies only weakly ($\propto n^{1/2}$).  The properties derived 
from the {\tt sedov} model fits should represent values averaged over the SNR.

The Sedov solution depends on three parameters: the explosion energy, 
$E_0$, the ambient mass density, $\rho_0$, and the time since the explosion,
$t_0$.  The ambient density of hydrogen, $n_0$, is related to the mass
density by $n_0 = \rho_0 / (1.43\,\mathrm{amu})$, where we assume
a fully ionized gas with half-solar abundances (based on the
\citealt{1989GeCoA..53..197A} solar abundance set).
The {\tt XSPEC sedov} model fits provide estimates for the postshock 
temperature, $T_\mathit{s}$, the ionization timescale, 
$\tau \equiv n_\mathit{e,s} t_0$, and also a model normalization 
$K = (10^{-14} / (4\pi D^2)) \int\,dV \,n_\mathit{e} \, n_\mathrm{H}$.
In the ionization timescale, $n_\mathit{e,s}$ is the postshock electron number
density (where we are assuming electron-ion equilibration in the shock),
and $n_\mathit{e,s} \approx 4.8\times n_0$ for our assumed abundances, and
assuming strong shock jump conditions.
For \oursnr{}, we also have an estimate for the physical radius of the SNR, 
$R_\mathit{s}$, based on the angular radius, $\theta_\mathit{s}$, and 
the distance to M33.  With four observational constraints and three model 
parameters, the system is overconstrained.  As noted by 
\citet{1998ApJ...505..732H}, the age estimates based on the Sedov dynamical 
age and the age based on ionization timescale allow for a consistency check. If 
the SNR is well modeled by a Sedov solution, these two age estimates should 
be the consistent with each other.  

The X-ray temperature implies a shock velocity,
\begin{equation}
\label{eqn:def.v.s}
v_\mathit{s} 
  = \left( { 16 k T_\mathit{s} \over 3 \mu_\mathit{s} } \right)^{1/2} 
  = 904\, T_{\mathit{s},\mathrm{keV}}^{1/2} \,\kms{}
\end{equation}
where $kT_{\mathit{s},\mathrm{keV}}$ is the postshock temperature
in \keV{} and $\mu_\mathit{s}$ is the postshock mean mass
per free particle.  The equation assumes a strong shock into a monatomic
gas with three degrees of freedom, and complete equilibration between
species \citep{1980ARA&A..18..219M}.  
For a fully ionized plasma with half-solar abundances, 
$\mu_\mathit{s} = (m_\mathrm{H} + A_\mathrm{He} m_\mathrm{He} + \cdots)/(2 +
3 A_\mathrm{He} + \cdots)
\approx 0.61$, where $m_\mathit{H}$ and $m_\mathit{He}$ are the atomic 
weights of H and He, respectively, and $A_\mathrm{He}$ is the 
fractional abundance of He relative to H.  From the {\tt sedov} 
model estimate for the postshock temperature, we have 
$v_\mathit{s} = 611\pm 12\kms{}$ where 
the error is a purely statistical error, and almost certainly underestimates the 
real uncertainty in the velocity.

If the SNR is assumed to be expanding as $R_\mathit{s} \propto
t^\delta$, the shock velocity, combined with the current SNR radius, 
allows for an estimate of the SNR dynamical age, $t_0$.  The model 
normalization, $K$, combined with $v_\mathit{s}$, provides an estimate 
for the ambient hydrogen density, $n_0$, and the postshock electron density, 
$n_\mathit{e,s} = 4.8\,n_0$, can be estimated by using the strong 
shock jump conditions and the (half-solar) abundances.  The postshock 
density can be combined with the ionization timescale estimate, 
$\tau$, from the {\tt sedov} model fit to provide a second estimate for 
the SNR age.  From these basic parameters, the swept up mass, 
$M_\mathit{SU}$, and the explosion energy, $E_0$, can be estimated.  We 
proceed to the evaluation of the parameters based on the {\tt sedov} fit 
parameters from simultaneous fits to the data for the whole SNR (the right 
hand column in Table~\ref{tbl:gkl21.fits}).

The mean angular radius of the X-ray SNR is $2.56\arcsec\pm 0.13\arcsec$, 
where the uncertainty in the radius includes a 2\% angular size uncertainty 
and 3.5\% uncertainty (added in quadrature) for the azimuthal asymmetry.  
This angular radius and the assumed distance to M33 
imply a radius of $10.1\pm 0.9 \pc{}$ for the SNR, where the distance 
uncertainty is now included.  For a Sedov-stage SNR, the deceleration parameter 
is $\delta=2/5$, and we estimate the dynamical age to be 
$(2/5) R_\mathit{s}/v_\mathit{s} \approx 6500\pm600\,\mathrm{yr}$.
Given the radial dependence of the electron and H number densities, the
emission integral, $\int\,dV \,n_\mathit{e}\,n_\mathrm{H}$, can be evaluated 
from the model normalization.  For the case of the Sedov solution, 
\begin{equation}
\int\,dV\,n_\mathit{e}\,n_\mathrm{H} \approx 4\xi
      \left(
          {n_\mathit{e} \over n_\mathrm{H} }
      \right)
      n_0^2 
      \left(
          {4\pi \over 3} R_\mathit{s}^3
      \right)
\end{equation}
where $n_0$ is the preshock H density and $\xi = 0.517773$ is an
integration constant evaluated by numerical integration over the Sedov
solution.  The average preshock H density can be estimated as
\begin{equation}
n_0 = 1.58 \, K^{1/2}_{-4}\, D_{800} \, R_\mathit{s,10}^{-3/2}
   \, \cm^{-3} ,
\end{equation}
where $K_{-4}$ is the {\tt XSPEC sedov} model normalization in units of $10^{-4}
\cm^{-5}$, $D_{800}$ is the distance to the SNR in units of 800\kpc{},
and $R_\mathit{s,10}$ is the radius of the SNR in units of 10\pc{}.  From 
the {\tt sedov} fit parameters, we obtain an average preshock H density of 
$n_0 = 1.7\pm0.3\,\cm^{-3}$.  Approximate pressure equilibrium around the 
rim of the SNR would imply that the temperature should vary as 
$\sim n_0^{-1}$.  Although the temperature derived from the fit to the 
BE region is somewhat higher than that obtained from the FW region fit 
(though not significantly so), even taking the full 90\% confidence ranges 
into account does not provide the $\sim2$:1 ratio implied by the density 
variation from the elliptical shell model fit (see 
\S\ref{sec:imaging.analysis}).  The relative overpressure of the BE 
region compared to the FW region may be an indication that the encounter 
with the density gradient toward the east is relatively recent.  This might 
also explain the lack of optical emission (see \S\ref{sec:imaging.analysis}) 
if there has not been sufficient time for the shocks to become radiative.
In principle, the ionization timescale provides a constraint on the age of the 
encounter; however, 90\% confidence intervals for $\tau$ for the BE and 
FW regions are broad enough to make this a weak constraint.
An additional possibility is that there are sufficient photoionizing photons
toward the east to prevent the postshock gas from recombining.  It is
notable that the edge of the enhanced \fion{S}{ii} in the east coincides
roughly with a bright rim of \Halpha{} emission associated with one of the
bright cores of the \HII{} region.

If a Sedov-stage SNR is assumed, the ionization timescale provides for another
estimate for the SNR age: 
$t_0 = \tau / n_\mathit{e,s} \approx 8200\pm 1700\,\mathrm{yr}$.  
The age based on ionization timescale is somewhat larger than the dynamical 
age, but the estimates are reasonably consistent.  The weighted mean of the 
estimates yields an age of $6700\pm600\,\mathrm{yr}$.

From the preshock H density and the radius of the SNR, the swept up mass 
is $\sim 260\pm80\,M_\sun$, assuming a constant preshock density (which
is almost certainly not the case, but would provide a reasonable estimate
if the density gradients are not too extreme).  For a Sedov-stage SNR, 
this mass, together with the postshock temperature, implies a total 
explosion energy of $E_0 \approx (1.8\pm0.3)\times 10^{51}\, \erg{}$.

\citet{1993ApJ...418..743G} estimated the nonthermal pressure
to be $P_\mathit{nt}/k \sim 2.4\times
10^{6}\,\mathrm{cm}^{-3}\,\mathrm{K}$
(where $k$ is the Boltzmann constant)
based on the radio flux and assuming a 28\pc\ diameter shell
with thickness $\Delta R/R = 1/12$ for the emitting region.
Scaling their pressure value to our smaller radius and using 
our Sedov model parameters to estimate the emitting volume, we 
obtain $P_\mathit{nt}/k \approx 4.6\times 10^{6}\,\mathrm{cm}^{-3}\,\mathrm{K}$.
As noted by \citealt{1993ApJ...418..743G}, the 
estimated nonthermal pressure is much smaller than the
thermal pressure estimate based on X-ray data 
(see Table~\ref{tbl:gkl21.cmp.snr}).
They note that the nonthermal pressure estimate is likely to be 
a lower limit:  pressure: field/particle nonequilibrium could increase the 
nonthermal pressure, and volume for the radio emitting plasma
(difficult to ascertain, because of the radio spatial resolution) 
could be much smaller than was assumed.

\subsection{Comparisons with Other SNRs}
\label{subsec:n49.comparison}

The asymmetric brightness distribution invites comparison with two 
LMC supernova SNRs which also show large variations in optical
and X-ray surface brightness: N49, and 0506-68.0 (also known as N23).  
Both SNRs show strongly asymmetric X-ray and optical surface 
brightnesses as well as indications of interaction with denser material.
Table~\ref{tbl:gkl21.cmp.snr} summarizes the X-ray properties and 
Sedov model parameters for the SNRs, based on 
\citet{1998ApJ...505..732H, 2006ApJ...645L.117H}.  

\citet{1993ApJ...418..743G} compared \oursnr{} to the LMC SNR N49;
in the latter SNR, both the optical and the X-ray emission are strongly 
enhanced in the southeast.  A comprehensive study by 
\citet{1992ApJ...394..158V} examined optical and UV data, and developed a 
self consistent model for N49.  The optical emission is coming from slow 
(40--270\kms{}) shocks into dense (20--940\,cm$^{-3}$) gas as the SNR 
interacts with a molecular cloud to the southeast, while the X-ray 
emission results from faster (730\kms{}) shocks into an intercloud medium 
($n_0\approx0.9$\,cm$^{-3}$).  A CO cloud to the southeast of the SNR 
\citep{1997ApJ...480..607B} would account for the high densities there.  
\citet{1992ApJ...394..158V} inferred an age of 5400\,yr based on 
Sedov dynamics.  

SNR\,0506-68.0 also shows strong asymmetries in its X-ray, optical,
and radio morphologies \citep[and references therein]{2006ApJ...645L.117H}.  
Such asymmetries 
suggest that an interaction is taking place between the SNR and denser gas.  
However, searches for an associated molecular cloud have not been successful; 
moreover, the low absorption (as implied by the low column density) 
argues against the presence of such a cloud \citep{1998ApJ...505..732H}.  
\citet{2006ApJ...645L.117H} noted that the open cluster HS114 
\citep{1966AJ.....71..363H} lies near the high density side of the SNR 
(only 2\,pc in projection, assuming a distance of 50\kpc{} to the LMC).
They found at least a factor of 10 variation in the ambient density
around the rim of the SNR, and also detect regions with enhanced 
abundances in parts of the SNR.  By assuming that the side of the
SNR which is expanding into a region of low density can be modeled 
approximately as a free Sedov expansion, \citet{2006ApJ...645L.117H} 
estimated an age of $\sim 4600\,\mathrm{yr}$.

\clearpage
\begin{deluxetable}{llllllllll}
\rotate
\tabletypesize{\small}
\tablewidth{0pt}
\tablecaption{
  Comparison of \oursnr with LMC remnants N49 and 0506-68.0.
  \label{tbl:gkl21.cmp.snr}
}
\tablehead{
     \colhead{SNR}\hspace{2cm}
   & \colhead{$R_\mathit{s}$}
   & \colhead{$L_\mathit{X}$\tablenotemark{a}}
   & \colhead{$T_\mathit{s}$}
   & \colhead{$n_0$}
   & \colhead{$v_s$\tablenotemark{b}}
   & \colhead{$P_s/k\,$\tablenotemark{c}}
   & \colhead{$t_0$}
   & \colhead{$E_0$}
   & \colhead{$M_\mathrm{SU}$} \\[0pt]
     
   & \colhead{(pc)}
   & \colhead{$(10^{36}\,\mathrm{erg}\,\mathrm{s}^{-1})$}
   & \colhead{$(10^{6}\,\Kelvin)$}
   & \colhead{$(\mathrm{cm}^{-3})$}
   & \colhead{$(\kms)$}
   & \colhead{$(10^{7}\,\mathrm{cm}^{-3}\,\Kelvin)$}
   & \colhead{$(10^3 \mathrm{yr})$}
   & \colhead{$(10^{51} \mathrm{erg})$}
   & \colhead{$(M_\odot)$}
}
\startdata
\oursnr\dotfill 
   &  10                             %
   &  12                             %
   &  5.3(0.2)                       %
   &  1.7                            %
   &  610                            %
   &  8.3                            %
   &  6.7                            %
   &  1.8                            %
   &  260            \\[0pt]         %
N49\tablenotemark{d}\dotfill         %
   &  8.2                            %
   &  6.3                            %
   &  6.7(0.1)                       %
   &  2.6                            %
   &  690                            %
   &  16                             %
   &  4.4                            %
   &  1.5                            %
   &  210            \\[0pt]         %
0506-68.0\tablenotemark{d}\dotfill   %
   &  6.7                            %
   &  2.5                            %
   &  6.2(0.1)                       %
   &  1.6                            %
   &  660                            %
   &  9.1                            %
   &  3.8                            %
   &  0.46                           %
   &  70             \\[0pt]         %
0506-68.0\tablenotemark{e}\dotfill   %
   &  12                             %
   &  \nodata                        %
   &  \nodata                        %
   &  0.25                           %
   &  \nodata                        %
   &  \nodata                        %
   &  4.6                            %
   &  \nodata                        %
   &  \nodata                        %
\enddata
\tablenotetext{a}{0.5--5.0\keV}
\tablenotetext{b}{From Eq.~\ref{eqn:def.v.s}}
\tablenotetext{c}{$P_\mathit{s}/k = 4\times (2.3 n_0) T_\mathit{s}$,
where the 4 accounts for the strong shock density jump, and 2.3
is the number of free particles per H in a fully ionized plasma}
\tablenotetext{d}{\citet{1998ApJ...505..732H}}
\tablenotetext{e}{\citet{2006ApJ...645L.117H}}
\end{deluxetable}

\clearpage 

\oursnr{} shows similar asymmetries in the X-ray emission, with
the rim toward the \HII{} region core being much brighter than the
rest of the SNR.  Despite the X-ray similarities with the other SNRs, 
the optical emission shows rather different properties.  N49 and 
SNR\,0506-68.0 both show optical brightening on the same side as 
the X-ray brightening.  In contrast, \oursnr{} does not show strong 
optical brightening near the X-ray bright eastern rim (though the situation 
is confused by stellar contamination), but instead shows an optical 
extension toward the southwest well beyond the faint X-ray rim.
The strong X-ray emission ahead of the optical emission in the east is 
puzzling.  As noted in \S\ref{subsec:sedov.analysis}, this may indicate a 
relatively recent encounter with denser gas for which photoionization
associated with the bright \HII{} region core maintains the ionization
at a high enough state that any \fion{S}{ii} emission is weak.  
The extension of the optical emission well beyond the X-ray rim in 
the southwest is suggestive of some sort of breakout into a lower density 
region which has subsequently driven slower (hence X-ray faint) 
radiative shocks into denser material.  In that case it is puzzling
that no X-ray emission is seen within the breakout.  Possibly the 
X-ray emitting plasma driving the optical shocks is tenuous enough 
that the X-ray emission is too faint to be seen even in this 
190\ks{} exposure.  In any case, the peculiarities of the X-ray and 
optical emission for \oursnr{} suggest an explosion in a complex environment.

\section{Conclusions}
\label{sec:conclusions}

We have presented the results of an analysis of the \chase{} project
data for \oursnr{}, an SNR embedded in the giant \HII{} region NGC\,592 
\citep{1993ApJ...418..743G}.  These data provide the first well-resolved
X-ray imagery of an SNR in M33.  The remnant is slightly elliptical, and
roughly $5\arcsec$ in diameter.  We fitted a three dimensional
elliptical shell model to the X-ray distribution, from which we
measured the X-ray center of the SNR to be at
RA=$01^\mathrm{h}33^\mathit{m}11\fs73$,
Dec=+30$\arcdeg38\arcmin41\farcs9$ (J2000), and 
the dimensions of the SNR shell in projection to be
$\sim 21.0\,\mathrm{pc} \times 19.6\,\mathrm{pc}$ at the assumed distance
to M33.  The \Chandra{} data show the emission to be asymmetric,
with the eastern rim (closest to the bright \HII{} region cores)
roughly $5$ times brighter than the rest of the SNR.  This asymmetry
suggests that the SNR is interacting with the \HII{} region and further
supports the argument that the SNR is actually embedded in,
rather than merely in projection against, the \HII{} region 
\citep{1993ApJ...418..743G}.
The association with NGC\,592 suggests that the progenitor could have
been a high mass star, and that the SNR was the result of a core collapse
supernova.

We augmented the high resolution X-ray data with other \chase{}
data taken further off-axis in order to improve the statistics for
X-ray spectral analyses.  By fitting with the {\tt XSPEC sedov} model,
we obtained an estimate for the shock temperature 
$kT = 0.46_{-0.02}^{+0.01}\,\keV{}$, an average preshock 
ISM (hydrogen) density of 
$n_\mathrm{0} \approx 1.7\pm0.3\,\cm^{-3}$, and an ionization
timescale $\tau \equiv n_\mathit{e,s} \,t_0 
 \approx 2.1_{-0.3}^{+0.2}\times 10^{12}\,\mathrm{cm}^{-3}\,\mathrm{s}$.
Although the imaging data show the SNR to be somewhat asymmetric,
the global average properties can be described approximately by a Sedov model
\citep{1999A&A...344..295H}.
The detailed properties would be expected to vary azimuthally around the
SNR because of the variation in the preshock density.  The surface 
brightness will vary most strongly with density ($\propto n^{2}$),
the postshock temperature varies $\propto n$, and the shock velocity
is relatively insensitive to the ambient density ($\propto n^{1/2})$.
However, fits to the high spatial resolution data for sectors 
containing the brightest emission (region BE) and the rest of the 
SNR (region FW) did not show any significant differences 
in spectral properties.

From the {\tt sedov} model estimates, we can also derive average 
values for other properties.  The expansion velocity of the SNR 
is $v_\mathit{s} \approx 611\pm12\,\kms{}$
based on the observed size of the remnant and the postshock temperature.
By assuming a Sedov expansion rate, we obtain a dynamical age estimate
of $6500\pm600\,\mathrm{yr}$.  The average postshock electron density,
$n_\mathrm{e, s} = 4.8\,n_0$, where $n_0 = 1.7\pm 0.3$, combined
with the ionization timescale, $\tau$, yields an ionization age
estimate of $8200\pm1700\,\mathrm{yr}$.  The estimates are reasonably 
consistent with each other, and the weighted mean of the ages
is $6700\pm 600\,\mathrm{yr}$.  The large amount of swept up 
mass, $260\pm80\,M_\odot$, also supports the argument that 
\oursnr{} is at least a middle aged SNR.  

Our fitted abundance value, $0.45_{-0.09}^{+0.12}$ solar, is 
consistent with the expected half-solar abundances appropriate to the 
galactocentric radius
of \oursnr{}.  Although clumps of enhanced ejecta have been found in 
other SNRs of a similar age (e.g., SNR\,0506-68.0 (N23); 
\citealt{2006ApJ...645L.117H}),
we do not find evidence here for enhanced abundances due to ejecta.
Fits to the SNR as a whole will dilute any ejecta signature in the
much larger mass of swept up ISM, while much smaller extraction regions,
BE and FW, yielded poor counting statistics for the available 190\,ks of
high resolution X-ray imaging data.

The total X-ray luminosity (0.25--4.5\,\keV{}) of the source is estimated to 
be $(1.2\pm0.2)\times 10^{37}\,\mathrm{erg}\,\mathrm{s}^{-1}$ (absorbed),
or
$(1.7\pm0.3)\times 10^{37}\,\mathrm{erg}\,\mathrm{s}^{-1}$
(unabsorbed),
in good agreement with the \XMM{} result of 
$(1.13\pm0.14)\times 10^{37}\,\erg{}\,\second^{-1}$ (absorbed),
based on the observations of \citet{2004A&A...426...11P}.
We searched for any emission from a point source or plerion.
We found no evidence for a significant excess of hard emission, and obtained
an upper limit of 
$L_\mathit{X} \le 1.2\times 10^{35}\,\mathrm{erg}\,\mathrm{s}^{-1}$
(2--8 keV) for the unabsorbed luminosity of any hard pointlike source if 
present.

We compared the X-ray images with available optical data
and found some puzzling morphological differences.  
In contrast to the slightly elliptical X-ray remnant, the optical remnant
is more oblong.  The bright optical emission agrees with the X-ray
contours in the north and northwest, but falls off more rapidly in the
east, with the region of brightest X-ray emission almost devoid
of optical emission (although this comparison is hampered by the
presence of several bright stars).  To the southwest, the
optical emission extends considerably beyond the X-ray
emission.  The lack of optical emission to accompany the bright X-ray
emission in the east could be the result of a relatively recent interaction 
with denser material, in which the shocked gas has not had time to
become radiative; the apparent overpressure in the BE region compared to 
the FW region would support that interpretation. The edge of the 
\fion{S}{ii} emission in the east coincides with the rim of an
\HII{} loop, and the bright X-ray emission may be the result of an interaction
with a bubble driven by the stars associated with the southern \HII{}
region core.  In that case, the lack of \fion{S}{ii} emission may be
the result of the photoionizing flux from the stars in the 
\HII{} region core preventing the postshock gas from recombining.
On the other hand, the lack of X-ray emission to accompany the 
optical emission in the southwest is also puzzling.  It may be that 
even with 190\,\ks{} of data, the X-ray emission in the southwest is too 
faint to be picked up.  The complex multiwavelength
morphology suggests that the shell is expanding into a non-uniform density
medium, as may be expected for an SNR evolving in an active star
forming region.

\acknowledgments

This work has made use of SAOImage 
DS9\footnote{\tt http://hea-www.harvard.edu/RD/}, developed 
by the Smithsonian Astrophysical Observatory (Joye \& Mandel 2003), 
the {\tt XSPEC}\footnote{\tt http://xspec.gsfc.nasa.gov/}
spectral fitting package (Arnaud 1996), the 
FUNTOOLS\footnote{\tt http://hea-www.harvard.edu/RD/funtools} 
utilities package,
the HEASARC 
FTOOLS\footnote{\tt 
http://heasarc.gsfc.nasa.gov/docs/software/lheasoft/ftools/}
package, 
and the 
CIAO\footnote{\tt http://cxc.harvard.edu/ciao/}
(\Chandra{} Interactive Analysis of Observations) package.
Support for this work was provided by the National Aeronautics and Space
Administration through \Chandra{} Award Number G06-7073A issued by the 
\Chandra{}
X-ray Observatory Center, which is operated by the Smithsonian
Astrophysical Observatory for and on behalf of the National Aeronautics
Space Administration under contract NAS8-03060.  TJG, PPP, and RJE 
acknowledge support under NASA contract NAS8-03060.

{\it Facilities:} \facility{CXO (ACIS)}

\end{document}